# Cyber Threat Intelligence for Secure Smart City

Najla Al-Taleb[1], Nazar Abbas Saqib[1], Atta-ur-Rahman[1], Sujata Dash[2]
College of Computer Science and Information Technology, Department of Computer Science
[1] Imam Abdulrahman Bin Faisal University, King Faisal Road, King Faysal University, Dammam, Saudi Arabia
[2] North Orissa University, Barispada, Odisha, India
2190500053@iau.edu.sa, nasaqib@iau.edu.sa , aaurrahman@iau.edu.sa, Sujata238dash@gmail.com

*Abstract*—Smart city improved the quality of life for the citizens by implementing information communication technology (ICT) such as the internet of things (IoT). Nevertheless, the smart city is a critical environment that needs to secure it is network and data from intrusions and attacks. This work proposes a hybrid deep learning (DL) model for cyber threat intelligence (CTI) to improve threats classification performance based on convolutional neural network (CNN) and quasi-recurrent neural network (QRNN). We use QRNN to provide a real-time threat classification model. The evaluation results of the proposed model compared to the state-of-the-art models show that the proposed model outperformed the other models. Therefore, it will help in classifying the smart city threats in a reasonable time.

*Keywords—Smart city, CTI, Threat, Security, Privacy, Attack, DL, CNN, QRNN*

## I. Introduction

The concept of a smart city is increasing over the world, where different cities such as Dubai, Barcelona, and New York start becoming more intelligent. These cities are providing services through technology such as IoT and Cyber-Physical Systems (CPS), where they are connected through a network to monitor, control and automate the city services to provide the best quality of life for the citizens [1].

Smart city technologies exchange and process different types of data to provide services. These data can be sensitive and critical which imposes security and privacy requirements. However, the characteristics of smart city technology such as IoT and CPS in terms of resources limitation such as power, memory, and processing imposes challenges to run sophisticated security mechanisms and expose smart city infrastructure to cyber-attacks [2]. Therefore, different attacks target smart city infrastructure including Distributed Denial of Service (DDoS) using IoT devices by infecting IoT devices by bots and lunch the attack against the attack target [3].

A CTI can provide a secure environment for smart cities, where it can rely on cloud services to monitor the possible threats in real-time in a smart city and take the appropriate prevention measures without human intervention. Moreover, it will provide light security mechanism to the smart city systems since it will not be implemented on smart city devices, rather than that, it will monitor the attacks in the smart city through the cloud to get information about the recent threat behavior and indicator of compromise (IoC) to report them to the connected smart city systems.

Different techniques and models are proposed to analyze cyber threats for CTI such as machine learning (ML) and deep learning (DL) models. Nevertheless, these artificial intelligence (AI)-based models can have a high false-positive rate (FPR) and low true-positive rate (TPR) if the attack traffic is not profiled well and not modeled enough [4]. Thus, it limits the real-time classification efficiency and degrades the smart city network security. Therefore, to address this issue and improve the threat analysis and lower the FPR, we propose a hybrid DL model that is based on convolutional neural network (CNN) and quasi-recurrent neural network (QRNN). The proposed model can learn spatial and temporal features automatically without human intervention. We evaluate our proposed model on two IoT network traffic datasets. The evaluation results demonstrate the effectiveness of our proposed model.

The rest of this paper is structured as follows. In Section 2, we discussed CTI and the concept of a smart city including security challenges and cyber threats in a smart city. In Section 3, we compare and analyze different threat classification schemes that are proposed in the literature. The proposed model is presented in Section 4. The implementation of the proposed model is provided in Section 5 and in Section 6 the experiment results and analysis are presented. We conclude the paper in Section 8.

## II. Background

*A. Smart City*

The smart city concept refers to urban systems that integrated with ICT to improve city services in terms of monitoring, management, and control to be more efficient and effective [5]. The smart city contains a huge number of sensors that continuously generate a tremendous amount of sensitive data such as location coordinates, credit card numbers, and medical records. These data are transmitted through the network to data centers for processing and analysis to take the appropriate decisions such as managing traffic and energy in a smart city [6][3].

Sensors that generate data and devices that handle the data in a smart city have vulnerabilities that can be exploited by cybercriminals. Consequently, the citizen's privacy can be at risk as well their life in case of data that to be processed for taking decisions is manipulated, which makes the people intimidated to move to the smart city [1].

In the rest of this section, we discuss the smart city services and it's security and privacy, where we discuss the security requirements and the threats that target smart city.

*1) Smart City Services and Components*

Aiming to provide a better life for the citizens, different smart city services have been developed considering different aspects such as energy, living, environment, and industry [7]. The smart city concept is very complex due to the different components that are connected in smart city to generate and exchange data and take decisions and actions. Based on [3], the components of smart city can be classified into the following categories.

- Smart city devices: There are different devices integrated in a smart city to collect different types of data such as

sensors, a programmable logic controller (PLC), and smart streetlights and CCTV [3].

- Smart city systems: Smart city systems include a database (DB), server, human-machine interface, and management system [3].
- Smart city network: Devices, systems and services in smart city communicate through smart city network using different protocols such as Zigbee, TCP/IP, and Wi-Fi [3].

Even though the smart city provides such convenient and intelligent smart applications, it is still facing some challenges that affect the functionality and connectivity of these applications such as resource constraints devices and heterogeneity in terms of data type, networks, platforms, etc. [7]. Thus, it is affecting the decision of selecting the appropriate security measures.

*2) Security and Privacy in Smart City*

Smart city environment collects a tremendous amount of private and sensitive data and depends on ICT which make smart city target for cyber-attacks [8]. Therefore, security requirements such as authentication, integrity, confidentiality, and authorization are not just features added to the systems, but it should be essentials that must be integrated into smart city components and services.

The security requirement in a smart city can be divided into two types operational security requirements and data security requirements. The operational security requirements safeguard the smart city infrastructure against cyber-attacks, while data security requirements depend mainly on the operational security requirements such as if the system is hacked, the attacker will be able to expose the data [6].

*a) Smart City Security Challenges*

Maintaining security in a smart city is a challenge due to different reasons such as the resource constraint devices and the heterogeneity in a smart city [9]. Some security challenges in a smart city are:

- Smart city interconnectivity: The interconnective of smart city processes can be defined as communication and sharing data between different parties and organizations. These organizations handle the data differently based on their policies and priorities [10]. Consequently, it imposes a challenge to set privacy and security standards to handle the data, since it can affect the functionality and productivity of the organizations.
- Heterogeneity: In a smart environment that is based on IoT-systems, heterogeneity imposing a challenge due to the variety of IoT nodes, communication protocols, platforms, and hardware performances [7]. Thus, it is difficult to set security and privacy standards that can be implemented uniformly.
- Resource constraints: The nature of IoT devices and other smart city devices in terms of resource constraints including limited memory, processing capabilities, and battery capacity imposes security challenges [6][7]. The resource constraints devices can't handle and implement sophisticated security algorithms, which make them vulnerable to cyber-attacks.

*b) Cyber Threats in Smart City*

According to [3], cyber threats in a smart city can be categorized into three types, based on the target, as cyber threats against devices, cyber threats against systems, and cyber threats against networks [3].

Although various techniques proposed to improve the cyber security of a smart city, the attack techniques are improving as well, which put the smart city vulnerable to contemporary cyber-attacks such as spam attacks, Sybil attacks, and identity attacks [7]. In addition to these threats, the following threats are some of the latest threats that target a smart city.

- IoT-Botnets: The IoT botnet is one of the serious cyber threats in a smart city, due to the nature of the botnets and the number of attacks that can be launched using IoT botnets. Botnets consist of multiple devices, called bots, that are connected through the Internet to implement a task repetitively while communicating with command and control (C&C) server to take instructions [11]. For example, Mirai botnet can infect devices such as webcams, routers, and IP cameras to build botnet and infect more IoT devices to lunch DDoS attacks against targeted servers [7].
- DDoS: In DoS attack, an attacker attempts to overwhelm a machine or network resource with a flood of requests by disrupting services of a host connected to the Internet to make them unavailable to its legitimate users. In terms of DDoS attack, the flood of requests on the targeted device is flooding from multiple sources, which make it difficult to track the sources and prevent the attack. For smart cities, many devices, such as parking meters and IoT sensors, can be attacked and breached to be controlled through IoT botnet that programmed to flood a system with requests simultaneously [3].
- Privacy and Identity Theft: Smart city infrastructure generates a huge amount of data related to provided services such as credit card data and surveillance feeds. The nature of the unprotected smart city infrastructure and the huge amount of personal information is appealing to the cyber attackers to exploit them and lunch identify theft to impersonate the target identity [11]. For example, the IoT botnet in a smart city can be used to data theft by recording the user keystrokes and steal their credentials information [12].
- Man-in-the-middle (MITM): An attacker secretly interrupts, spoofs, or alters the connection between two systems, by obtaining the authentication information and masquerading as a legitimate user [3]. For example, the attacker deliberately eavesdrops or alter messages between two smart vehicles in a smart city which is too risky [13].
- Device hijacking: The attacker hijacks a device and undertakes control of it. In many cases of these attacks, the attacker does not manipulate the basic functionality of the device. Thus, it can be difficult to detect device hijacking. For a smart city, an attacker could hijack and take control of autonomous vehicles in a smart city to change the predefined routes [14].

*B. Cyber threat intelligence (CTI)*

In recent years, the popularity of CTI increased in information security, especially in the area of IoT which are widely used in different areas including smart cities, due to the CTI effectiveness in analyzing different types of threats including APTs [15]. To understand the concept of CTI, it is required to know what intelligence is. According to [16], the term intelligence can be defined as the actionable format of the information, which is collected data from the environment that is processed to produce information. Therefore, threat intelligence can be defined as the collected data, which is related to security threats, malware, IoC, and vulnerabilities, processed to take actions [17][18].

The CTI helped in the security field to analyze the emerging threats to take prevention measures by analyzing threat data to identify the IoC for cyber-attacks, understand the behavior of the threat actors and their TTP, and the behavior of cyber-attacks in the network. As a consequence, the security system or the analyst can take the correct prevention measure in time [17]. One of the aspects that is important for CTI is the automation of CTI, where automated sharing and analyzing cyber threat data is required to integrate CTI in different fields such as smart cities [11]. The CTI can be implemented in cloud or any base knowledge that capable of processing and handling a huge amount of data with complex algorithms due to the significant amount of threat data that are collected from different security monitoring systems [17].

*1) CTI Data*

Since CTI can be described as evidence-based knowledge, the data that are collected and analyzed considered as the most important object in CTI because the action will be taken based on the analyzed data. For example, IoC can be used in threat hunting and to update the signature-based attack detection system [18]. The cyber threat data can be categorized into the following types of data.

*a) Low-level cyber threat data*

The low-level threat data or low-level IoC includes IPs, network artifacts, hashes, keystroke, windows event log, and more which are the most used cyber threat data in CTI and IDS [19][20]. The low-level threat data showed an effective threat data analysis due to the low-level data threat pattern which helps in identifying and profiling the threats [19]. One of the disadvantages of this type of data that it is atomic in nature, where the threat actor can change these IoC dynamically to evade and bypass the detection and prevention techniques [19][21].

*b) High-level cyber threat data*

The high-level threat data includes attacker behavior such as TTP, motivation, and attack patterns. This type of data can help in identifying the threat actor technique and link it with the attack to identify the attack motivation. As a consequence, the system will be able to take the appropriate measures to prevent this attack [19].

One of the disadvantages of this type is the need to involve humans in extracting knowledge. Due to the textual and unstructured type of threat data, the need of selecting the appropriate and accurate type of information or keywords that should be fed into the machine to extract the targeted knowledge from different sources is a challenge. The selection of the keywords can affect the knowledge extraction from different sources by discarding important and critical data. Consequently, it will affect the threat analysis and profiling threats and thereat actor [21].

These types of cyber threat data are provided by different sources, that insure to provide the threats relevant information on time such as FireEye, IBM X-Force, and Threat Tracer [18].

*2) CTI Techniques*

Different AI techniques used in CTI either to extract the knowledge from raw data or in data analysis. AI showed its usefulness at most in CTI by providing high performance in detection and profiling rates which helped in cyber security field to deal with the emerging types of threats and to mitigate the possible loss in organizations [22].

NLP is one of the areas in AI and it is used in CTI to extract data from unstructured textual data, by identifying the targeted topics that are provided to the NLP model to construct structured data [22]. Various CTI used NLP to extract data from threat blog reports, the dark web and other CTI reports. The following techniques are subcategories of AI that are used to analyze data in CTI.

In ML models, the algorithm trained and programmed to do a specific task. Different ML models used in cyber security for different functions such as to detect attack patterns [23]. Feature selection process in ML is required to improve the performance of the model [24]. Support vector machine (SVM) is one of the most used ML models and it is used to solve an optimization problem in finding a decision boundary to classify the data [25]. However, with the new threat actor techniques and the complex cyber-attacks, the traditional ML models are unable to detect these attacks [26].

DL is a subcategory of ML, that is more complex than ML. In the learning process in DL model, the algorithms attempt to learn from data at various levels according to the different levels of abstraction. Thus, a DL model performed better than ML model in terms of the feature extraction process. There are different architectures of DL including hybrid DL, supervised, and unsupervised [27].

III. LITERATURE REVIEW

In recent years, different papers proposed mechanisms to predict and analyze cyber-attacks and to provide security in a smart city especially privacy threats such as information theft and reconnaissance.

The authors in [28], proposed an ML-based detection mechanism that focused on classifying DDoS patterns to protect the smart city from these threats. However, they didn't consider the smart city environment and needs such as FPR and the time for identifying the pattern to take proactive measures in the connected systems. In [29], the authors studied how can IoT devices affect smart city cyber security. The authors proposed a detection mechanism that depends on the features that to be selected to improve the detection for IoT, to classify each type of threat. The results of the proposed system showed high accuracy, but the dataset, KDD CUP 99, doesn't represent the behavior of IoT network attacks.

Soe et al. [30] proposed an algorithm to select the best features for each type of attack in IoT environment to improve prediction accuracy and to provide a lightweight detection system. The authors used ML models to evaluate the proposed feature selection algorithm, which was able to predict the threats accurately. However, the proposed algorithm selects a static set of features for each type of attack, which can be easily bypassed if it is exposed to the threat actor. While in [31], the authors used DL model to classify cyber threats in IoT environment. The authors addressed the use of DL model to select the best features for threat prediction to improve the detection time. The proposed model selects a set of features, which are fed into feed-forward neural networks (FFNN) to detect cyber threats and classify the type of threat. However, the proposed model showed low accuracy in predicting information theft data.

In [32], the authors discussed how to use ML model to rapidly and efficiently detect and classify IoT network attacks. The authors performed an experimental study in which they implemented various ML models to evaluate their performance. While in [33], the authors proposed a hybrid ML model to detect IoT network attacks including the zero-day. The proposed model consists mainly of two stages, the first stages classifies the traffic into normal or attack and the second stage classifies the type of the attacks using SVM. Similarly in [34], the authors proposed a hybrid ML model to detect and classify the IoT network attacks on time. The first layer of the proposed model uses decision tree classifier to detect malicious behavior then the second layer classifies the type of this attack using RF. While in [14], the authors investigated the remote control threat of the connected cars and used an ML model to predict threats. The authors addressed the problem of real-time cyber threat detection and how the existing solution is based on reactive signature detection. Therefore, the authors proposed a proactive anomaly detection that profiles the behavior of the autonomous connected cars using Recursive Bayesian estimator. To evaluate the proposed method, the authors designed a dataset for connected cars using events routs, that are not real, and the global positioning system coordinates, then model the data analysis to predict the anomalies' behavior.

Lee et al. [23] proposed a technique, based on DL models, that transforms the multitude of security events to individual event profiles. The authors discussed how anomaly-based detection can be costly since it will trigger many false alerts. Therefore, they focused on improving the security information and event management system by using DL to reduce the cost to differentiate between true and false alerts. In [35], the authors proposed a hybrid ML method to detect cyber threats. The author focused on how to improve the detection accuracy to handle the attacker's methods to evade the detection tools. To evaluate the proposed method the authors used different datasets including KDD Cup and UNSW-NB15. While in [36], the authors discussed how to improve the threat analysis and classification including novel attacks. The authors proposed a model based on stacked autoencoder to enhance and automate feature selection to classify the threats.

Various works proposed a hybrid DL model to improve threat analysis and classification. In [37], the authors proposed an improved version of grey wolf optimization (GWO) and CNN. In the proposed hybrid model, the first model GWO is used to select the features, and the second model CNN is used for threat classification. Different works used the hybrid DL model that is based on CNN and RNN for spatial and temporal feature extraction to improve attack classification. In [38], the authors used CNN for feature selection, since it will provide fast feature selection to support real-time analysis. For threat classification, the authors used one of the variants of the LSTM model which is weight-dropped LSTM (WDLSTM). The proposed hybrid model showed high performance in terms of execution time.

Vinayakumar et al. [39] studied the effect of CNN in threat classification and IDS. The authors investigated different hybrid DL models with CNN including CNN-LSTM, CNN-GRU, and CNN-RNN. The model that consists of CNN-LSTM outperformed the other models. Moreover, the authors highlighted that selecting a minimum set of features for threat classification degrades the performance of the classification. Therefore, DL models can perform better in terms of feature selection. In [40], the authors proposed a hierarchical model that is based on CNN-LSTM. The authors used stacked CNN layers for spatial features learning using image classification, then stacked LSTM for temporal features learning. Similarly, in [4] the authors proposed the LuNet model which is based on CNN-LSTM. The authors discussed how stacking LSTM layers after CNN layers can drop some of the temporal features. Thus, the authors proposed the LuNet block which consists of LSTM layer stacked after the CNN layer, then stack the LuNet block in multiple layers to improve the classification performance and lower the FPR.

*A. Comparison and analysis*

As shown in Table I, the low-level IoCs that are collected from the network traffic have been used to classify the attack type in various papers.

Different network traffic benchmark datasets used to analyze the low-level IoC such as UNSW-NB15, NSL-KDD, and KDD CUP 99. For IoT attack classification, the BoT-IoT dataset is used in multiple works to evaluate the performance of the proposed models. Different ML and DL models are used to analyze the threats and to provide accurate results such as SVM, CNN, and LSTM, where CNN-LSTM hybrid model is used in multiple papers to improve threats classification performance.

In terms of the CTI for smart city, multiple papers, including [28] and [29], analyzed the threats pattern based on the network traffic. DDoS is one of the challenging threats in a smart city that been studied in different papers where these works proposed methods either to analyze the IP address and track the sources to prevent this attack or to identify the behavior of the network when there is overload traffic. Data theft which can be described as privacy and identity theft is another threat that been studied in various works. Data theft threats include reconnaissance, information theft, Probe, R2L, and U2R which may lead to exposing various vulnerabilities that help in lunching data theft attacks such as sniffing passwords and unauthorized access.

Some of the proposed models for the smart city set a fixed threshold to detect attacks which is not an effective way and it can raise a lot of false alarms which will affect the power consumption of the connected system. Since in smart city the normal behavior of the system can change due to the increasing number of connected devices such as IoT. In terms of accuracy, some papers achieved high accuracy, but they didn't show good performance in terms of FPR.

*B. Discussion*

Even though different papers proposed models to enhance threats classification for IoT environment, they are still lacking in terms of one or more of the following challenges:

- Real-time classification: One of the limitations that is common between different methods is the performance time. The low-level IoCs that are collected from the network traffic have been used to analyze the threats in various papers, where it should give timely information to the CTI knowledge base to update the detection and prevention information for all the connected systems to the CTI. However, to enhance the classification performance, various models stacked multiple ML model layers.

TABLE I.    COMPARISON BETWEEN THE PROPOSED ATTACK CLASSIFICATION METHODS

| Ref | Cyber Threats | Algorithm | Data Sources | Accuracy | FPR |
|---|---|---|---|---|---|
| [28] | DDoS | Restricted Boltzmann machine and FFNN | Simulated smart water system dataset | 97.5% | - |
| [30] | Information theft, reconnaissance, and DDoS | J48 | BoT-IoT UNSW | - | 0.41 |
| [31] | Information theft, reconnaissance, and DDoS | FFNN | BoT-IoT UNSW | - | - |
| [32] | DDoS, DoS, data exfiltration, keylogging, OS fingerprinting and service scan | KNN | BoT-IoT UNSW | 99.00% | - |
| [33] | DDoS, DoS, keylogging, and reconnaissance | C5-SVM | BoT-IoT UNSW | 99.97% | 0.001 |
| [34] | DDoS, DoS, data exfiltration, keylogging, OS fingerprinting and service scan | Decision tree-RF | BoT-IoT UNSW | 99.80% | - |
| [14] | Remote car control | Recursive Bayesian estimation | Routes data for connected cars | - | - |
| [23] | DoS, Probe, R2L and U2R | FCNN, CNN, and LSTM | Network events | 94.7% | 0.049 |
| [35] | Tor traffic (anonymous IP) | C4.5, MLP, SVM and LDA | UNB-CIC TOR Network Traffic dataset | 100 | 0 |
|  | Worms, DoS, backdoors, reconnaissance and more |  | UNSW-NB15 | 97.84% | 0.23 |
| [36] | Injection, Flooding, Impersonation | SAE | AWID-CLS-R | 98.66% | - |
| [37] | DoS, Probe, R2L and U2R | GWO-CNN | DARPA1998 | 97.92% | 3.60 |
|  |  |  | KDD CUP 99 | 98.42% | 2.22 |
| [38] | Worms, DoS, backdoors, reconnaissance and more | CNN-LSTM | UNSW-NB15 | 98.43% | - |
| [39] | DoS, Probe, R2L and U2R | CNN-LSTM | KDD CUP 99 | 98.7% | 0.005 |
| [40] | DoS, Probe, R2L, U2R, BruteForce SSH, DDoS and Infiltrating | CNN-LSTM | ISCX2012 | 99.69% | 0.22 |
|  |  |  | DARPA1998 | 99.68% | 0.07 |
| [4] | Worms, DoS, backdoors, reconnaissance and more | CNN-LSTM | UNSW-NB15 | 84.98% | 1.89 |
|  |  |  | NSL-KDD | 99.05% | 0.65 |

Therefore, it may take time to train the model and to classify the threats, thus it doesn't take advantage of these IoCs.

- FPR: When some models are not provided with a sufficient amount of data for each type of threat, the threat traffic will not be profiled well and modeled enough. Consequently, the ML models can have high FPR which is one of the challenges.

- Accuracy: As we said above, some models only give accurate results when the system has precise details of threats artifact. As a consequence, the system will not be able to recognize the threats that don't have a sufficient amount of data for model training which affects the classification accuracy.

In this work, we are proposing a hybrid DL model for CTI for smart cities that address the above challenges. The hybrid DL model can improve the threat classification accuracy as well as lowering the FPR in a reasonable time. Therefore, it can predict different attacks to protect the citizen's data and enhance the security of a smart city.

## IV. THE PROPOSED MODEL

In this section, we discuss the proposed hybrid DL model in terms of the structure, the selected DL algorithms, and their theoretical concepts.

Our proposed DL model consists of CNN and QRNN models. The selected DL models used to classify the threat type in real-time while providing a low FPR. The architecture of the proposed model is presented in Fig. 1. The benefits of using CNN and QRNN that they can increase the speed of the threat analysis while improving the accuracy of classification. In the rest of this section, we discuss the theoretical concepts of the proposed model.

### A. Convolutional Neural Network (CNN)

CNN is an extension of neural network and it proposed by Y. Le Cun et al. [41] and it's effective in extracting features at a low level from the source data, especially the spatial features [42]. CNN is used widely in image processing due to it is ability in automating feature extraction [43]. Also, CNN demonstrated its effectiveness in many fields such as biomedical text analysis and malware classification [23].

Based on the shape of the input data, the CNN can be classified into different types including 2-dimension (2D) CNN, which takes data such as images, and 1-dimension (1D) CNN, which takes data such as text data. The CNN consists of convolution layer, pooling layer, fully connected (FC) layer, and activation function [44]. The convolution layer is a fundamental building block in CNN that takes two sets of information as inputs and performs a mathematical operation upon these inputs. The two sets of information are the data and a filter, which can be called kernel. The filter is applied upon the entire dataset to produce the feature map [43]

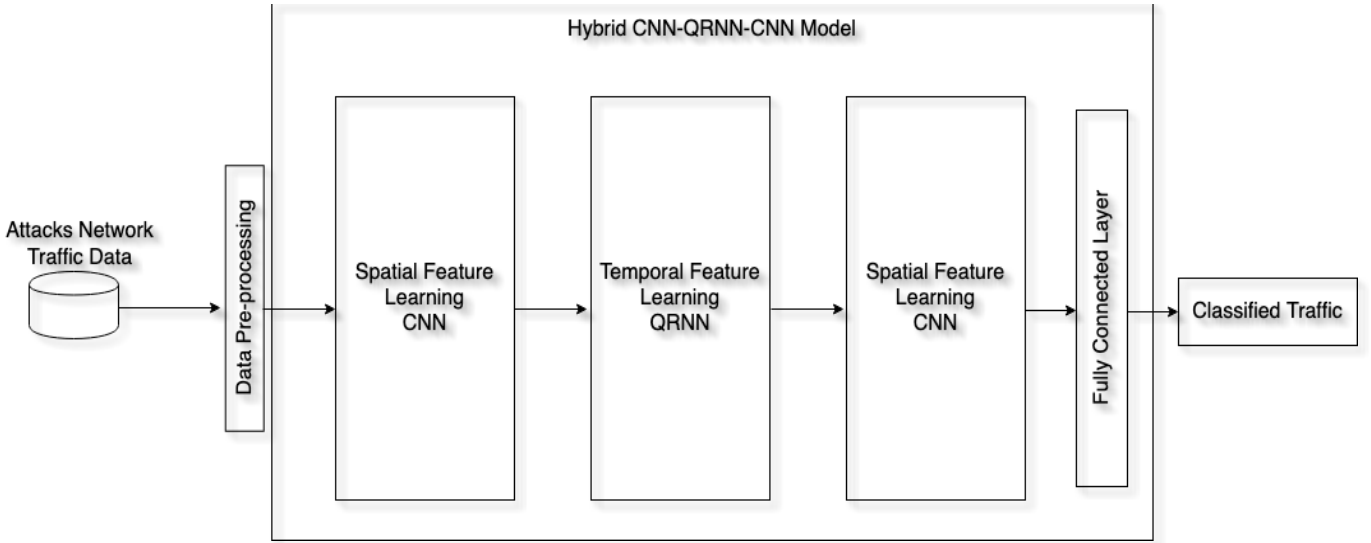

Fig. 1. The architecture of the proposed hybrid model

Each CNN filter extracts a set of features, feature map, which will be aggregated to a new feature map as output [23]. The pooling layer is implemented to reduce the feature map dimension and to remove irrelevant data to improve the learning [4]. The output of the pooling layer is fed into the FC layer to classify the data [45].

*B. Quasi-Recurrent Neural Network (QRNN)*

The LSTM-RNN is one of the most powerful neural network models that is used in cyber security due to its ability to accurately model the temporal sequences and their long-term dependencies [46]. However, LSTM usually takes a long time for model training and high computation cost [47]. In [48], QRNN is proposed. The QRNN model is designed to overcome the RNN limitation in terms of timestep's computation dependency on the previous timestep which limits the power of parallelism. The QRNN combines the benefits of CNN and RNN by using convolutional filters on the input data and allow the long-term sequence dependency to store the data of previous time steps [48]. The computation structure of QRNN is presented in Fig. 2.

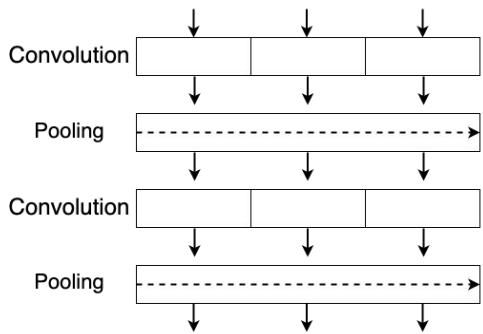

Fig. 2. The computation structure of QRNN

QRNN consists of convolutional layers and recurrent pooling function, which allow QRNN to work faster than LSTM by recording a 16 times increase in speed while achieving the same accuracy as LSTM in [49]. The convolutional and the pooling layers allow the parallel computation over the batch and feature dimension [48]. QRNN is used in different applications such as video classification [47], speech synthesis [49], and natural language processing [50].

*C. The proposed hybrid DL model for CTI*

Our hybrid DL model consists of 1D convolutional layer, 1D max pooling layer, QRNN, and FC layers. The first 1D convolutional layer selects the spatial features and produces a feature map that will be processed by the activation function. The Rectified Linear Unit (ReLU) activation function is used in the convolutional layers because of it is rapid convergence of gradient descent, which makes it a good choice for our proposed model [43]. The feature map then will be processed by the second layer, the pooling layer which we used the max pooling operation. The max pooling operation will select the maximum value in the pooling operation [43]. The pooling layer will reduce the dimensionality and remove the irrelevant features. The output of the CNN model will retain the temporal feature which will be extracted by the QRNN model.

Fig. 3 provides the details of our proposed model. We used two layers of QRNN to extract the temporal features. In the two layers of QRNN, the hidden size represents the number of the hidden units which represents the output dimension as well. The hidden units can be selected to the value of the number of features or above [47]. One of the problems of the neural network is overfitting, which means the model learns the data too well. Consequently, the model will not be able to identify variants in the new data [51]. Thus, we added a dropout layer to prevent overfitting.

Then, we used 1D convolutional layer and max pooling layer to extract more spatial-temporal features. The output of the CNN model is passed to the Flatten layer, which is a fully connected input layer, that transforms the output of the pooling layer into one vector to be an input for the next layer [52]. Finally, the dense layer, which is also a fully connected layer, with the softmax activation function is used to classify the threats by calculating the probabilities for each class [38].

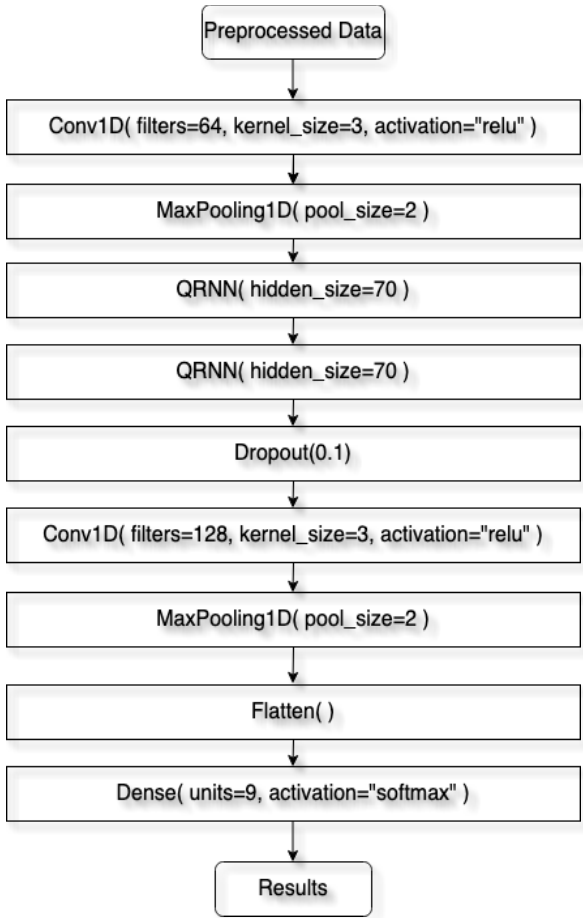

Fig. 3. Illustration of the details of the proposed model

## V. IMPLEMENTATION

In this section, we describe the datasets that we selected to evaluate the proposed model. Also, we discuss the data preprocessing steps, model parameter selection process, and the selected evaluation metrics.

### A. Datasets

In this work, we selected the following datasets because they are simulated to represent a realistic IoT environment such as smart home and smart city in terms of different aspects including:

- The heterogeneity of the simulated IoT devices including a weather monitoring system, smart lights, and smart thermostat.
- Different botnet scenarios such as probing and DoS attacks.

#### 1) Bot-IoT Dataset

Different datasets are used to evaluate ML models such as KDD99, ISCX, and CICIDS2017. Also, few datasets were produced to reflect a realistic IoT network traffic to evaluate ML models for IoT environment. However, these datasets were still lacking either the dataset is not diverse enough in terms of the attacks or the testbed is not realistic [32]. Thus, Moustafa et al. [12] designed the BoT-IoT dataset to address these limitations.

The Bot-IoT dataset is used in forensic analysis and to evaluate IDS. The dataset contains normal IoT traffic and different types of attack traffic with subcategories for each type of attack which are listed in Table II. Information gathering or reconnaissance is one of the privacy threats and it allows the threat actor to collect data about the victim such as port scanning and OS fingerprinting. While information theft includes data theft by unauthorized access to the data to download the data and keylogging. On the other hand, DoS threat affect the availability of the services and it can damage systems which makes it one of the biggest threats on smart city. UDP, TCP, and HTTP protocols were used to perform both DoS and DDoS attacks.

TABLE II. ATTACK CATEGORIES IN BOT-IOT DATASET

| Attack | Attack Subcategory | Number of Instances |
|---|---|---|
| Reconnaissance | Service scan | 73168 |
| | OS fingerprinting | 17914 |
| DoS | TCP | 615800 |
| | UDP | 1032975 |
| | HTTP | 1485 |
| DDoS | TCP | 977380 |
| | UDP | 948255 |
| | HTTP | 989 |
| Information theft | Keylogging | 73 |
| | Data theft | 6 |

The dataset was generated in the center of UNSW Canberra Cyber by using a testbed that consists of the three elements which are network platforms using different virtual machines, IoT services that are simulated using the Node-red tool, which contains different IoT services such as weather station and extracting features and forensics analytics. We used the train-test BoT-IoT dataset to evaluate our proposed model.

#### 2) TON_IoT Dataset

ToN_IoT dataset [53] is one of the newest cyber security datasets that was generated by Dr. Nour Moustafa at the Cyber Range and IoT Labs of the UNSW Canberra Cyber. The dataset was collected from a testbed network for industry 4.0 IoT and Industrial IoT (IIoT) which makes it suitable to evaluate CTI for a smart city.

We used the TON_IoT train-test dataset, which is in CSV format. The dataset contains a total of 461043 instances and 9 types of attacks which are presented in Table III with the number of instances for each type.

TABLE III. ATTACK CATEGORIES IN TON_IOT DATASET

| Attack | Number of instances |
|---|---|
| DoS | 20000 |
| DDoS | 20000 |
| Scanning | 20000 |
| Ransomware | 20000 |
| Backdoor | 20000 |
| Injection | 20000 |
| Cross-site Scripting (XSS) | 20000 |
| Password | 20000 |
| Man-In-The-Middle (MITM) | 1043 |

## B. Data Preprocessing

*1) Delete normal traffic records:* Since we are evaluating a CTI for threat classification, we deleted the normal traffic from the datasets. Also, in the Bot-IoT dataset, we have omitted the pkSeqID feature since it represents an identifier for the traffic records.

*2) Convert categorical features:* The datasets contain some categorical features that can't be processed by the neural network. Thus, we converted the nominal values into numeric using sklearn LabelEncoder. LabelEncoder converts categorical values into numerical values.

*3) Data standardization:* Many ML models may perform poorly on datasets with high data distribution. Thus, it will affect the learning efficiency of the model [51]. We implemented sklearn StandardScaler to scale the data.

*4) Split the data into training and testing:* For training and evaluation, we divided the data into training and testing datasets with a ratio of 35% for testing while considering having the same ratio of classes in both parts by using the stratify parameter.

## C. Model Implementation

The parameters of the hybrid model obtained during the training phase by trial and error including the number of CNN filters, number of QRNN hidden units, and dropout rate. For the kernel size, the values 3 and 5 are the most common values and the kernel size 3 performs well in this work with both datasets [39]. the filter size can help in extracting more details from the dataset by increasing the number of filters [54]. Thus, for the first CNN layer, we used 64 filters and for the other CNN we used 128 filters. The details and the selected parameters of the hybrid DL model are presented in Fig. 3.

## D. Evaluation tools and Metrics

To evaluate the ML models, it is important to select the appropriate evaluation metrics. Different evaluation metrics are used in this work to evaluate the performance of the proposed model including accuracy, FPR, true positive rate (TPR), precision, recall, and F-Score. Accuracy represents the ration of the correctly classified threats to the total number of classified threats. FPR represents the ratio of misclassified data as a different type of threat. While TPR represents the model ability to correctly classify the threats. Precision, recall, and F-Score are used to evaluate the overall performance of the proposed model, where a high value of precision indicates a low FPR. While recall represents the model's ability to correctly classify threats. The following equations represent the evaluation metrics where FP is false positive, TP is true positive, TN is true negative, and FN is false negative.

$$\text{Accuracy} = \frac{TP + TN}{TP + TN + FP + FN}$$

$$\text{FPR} = \frac{FP}{FP + TN}$$

$$\text{TPR} = \frac{TP}{TP + FN}$$

$$\text{Precision} = \frac{TP}{TP + FP}$$

$$\text{Recall} = \frac{TP}{TP + FN}$$

$$\text{F-Score} = \frac{2(\text{Precision} * \text{Recall})}{\text{Precision} + \text{Recall}}$$

## VI. EVALUATION AND ANALYSIS

This section presents the results and the analysis for model implementation. For implementation and evaluation, we used Jupyter Notebook software with Python programming language. We used Keras and scikitlearn packages for data pre-processing and to implement the proposed model. We trained the proposed model on a MacBook Air with Intel Core i5 CPU 1.6 GHz processor and 8 GB RAM. Also, we implemented different state-of-the-art ML models on the datasets to compare their performance with our proposed model.

Fig. 4 presents the confusion matrix of using our proposed model on the BoT-IoT dataset. The figure shows that the model correctly classified most of the cyber threat categories. Furthermore, to illustrate the quality of the proposed model, the receiver operating characteristic (ROC) curve is plotted in Fig. 5 for Bot-IoT dataset.

Fig. 6 presents the confusion matrix of using our proposed model on TON_IoT dataset, and the ROC curve is presented in Fig. 7 for TON_IoT dataset. In both ROC curves, our proposed model achieved the highest value which is 1. Thus, our proposed model performed very well with all the classes.

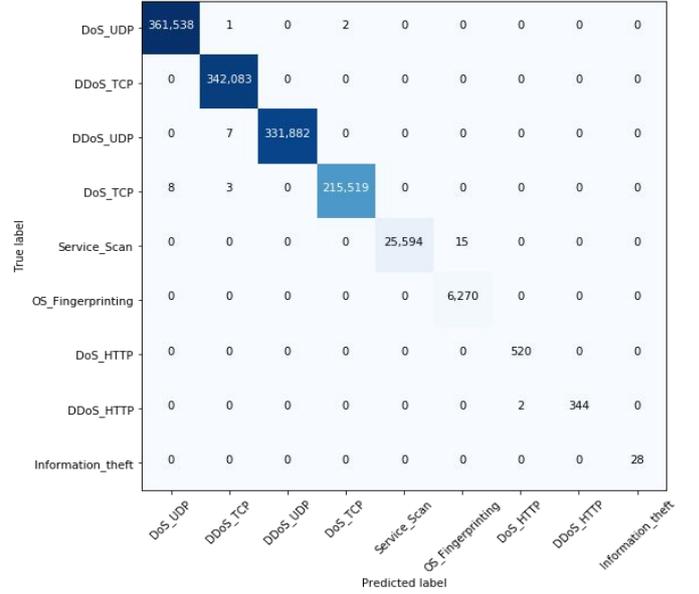

Fig. 4. The confusion matrix based on the Bot-IoT dataset

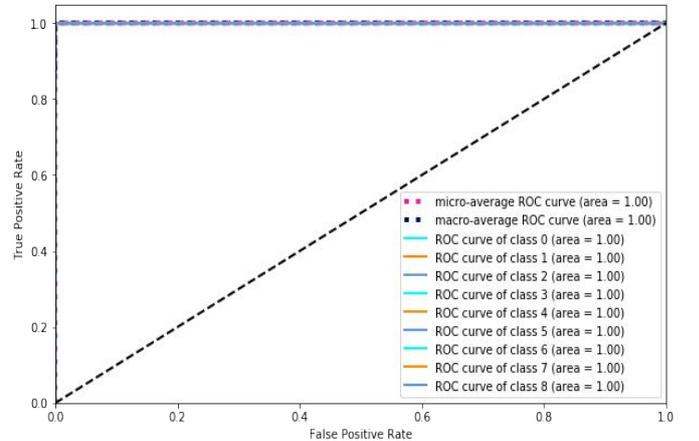

Fig. 5. ROC curve of using our proposed model on the Bot-IoT dataset

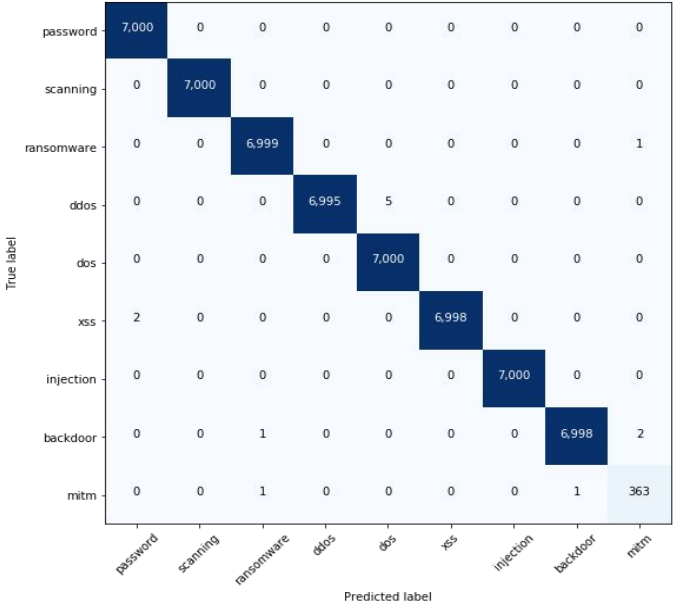

Fig. 6. The confusion matrix based on the TON_IoT dataset

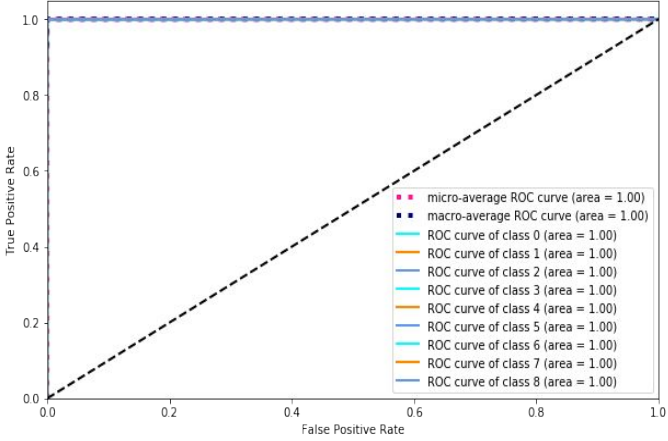

Fig. 7. ROC curve of using our proposed model on the TON_IoT dataset

The results of our proposed model on the testing datasets are presented in Table IV.

TABLE IV. RESULT OF CYBER THREAT CLASSIFICATION ON BOTH DATASETS

| Dataset | Accuracy% | TPR% | FPR |
|---|---|---|---|
| BoT-IoT | 99.99 | 99.92 | 0.0003 |
| TON_IoT | 99.99 | 99.99 | 0.001 |

As shown in Table IV, the proposed model achieves high accuracy with an average of 99.99% on both datasets. As well as the TPR, which reached an average of 99.92% with BoT-IoT dataset and 99.99% with TON_IoT dataset. Regarding the FPR, the proposed model achieved a low FPR with 0.0003 with BoT-IoT dataset and 0.001 with TON_IoT dataset. Thus, the proposed model showed a good performance in classifying the threats with both datasets.

Moreover, to demonstrate the effectiveness of QRNN in our model, we implemented our proposed model with LSTM instead of QRNN to compare the performance. The results are presented in Table V and Table VI.

TABLE V. COMPARISON OF OUR PROPOSED MODEL WHILE USING LSTM AND QRNN BASED ON BOT-IOT DATASET

| Model | Accuracy | Precision | Recall | F-Score | Avg. training time per epoch | Classification time |
|---|---|---|---|---|---|---|
| With LSTM | 99.99% | 100% | 100% | 100% | 1717.4 sec | 326 sec |
| With QRNN | 99.99% | 100% | 100% | 100% | 1299.1 sec | 251 sec |

TABLE VI. COMPARISON OF OUR PROPOSED MODEL WHILE USING LSTM AND QRNN BASED ON TON_IOT DATASET

| Model | Accuracy | Precision | Recall | F-Score | Avg. training time per epoch | Classification time |
|---|---|---|---|---|---|---|
| With LSTM | 99.99% | 100% | 100% | 100% | 86.3 sec | 16 sec |
| With QRNN | 99.99% | 100% | 100% | 100% | 66.5 sec | 13 sec |

Based on the results in Table V and Table VI, our proposed model with QRNN showed the same performance compared to our proposed model with LSTM in terms of accuracy, precision, recall, and F-Score. In terms of the time, the proposed model with QRNN showed a better performance for training the model and testing. Thus, QRNN showed its effectiveness in increasing the speed of the model while providing high accuracy and low FPR. Therefore, the model can be used for real-time CTI.

We compared the performance of our proposed model on Bot-IoT dataset against the state-of-the-art models for threats multi-class classification. The comparison is shown in Table VII.

TABLE VII. COMPARISON OF OUR PROPOSED MODEL WITH STATE-OF-THE-ART MODELS FOR MULTICLASSIFICATION BASED ON BOT-IOT DATASET

| Reference | Year | Accuracy% | Precision% | Recall% | F-Score% |
|---|---|---|---|---|---|
| [32] | 2019 | 99.00 | 99.00 | 99.00 | 99.00 |
| [33] | 2019 | 99.97 | - | - | 95.7 |
| [34] | 2020 | 99.80 | 99.00 | 99.00 | 98.80 |
| Our model | 2020 | 99.99 | 100 | 100 | 100 |

As shown in Table VII, our proposed model outperformed the other state-of-the-art models. Also, we implemented different ML models to compare their performance with our model. The accuracy, TPR, and FPR of each model with our model is given are Table VIII and Table IX. Our model performed better than the other four models due to the combination of CNN with QRNN.

TABLE VIII. COMPARISON OF OUR PROPOSED MODEL WITH OTHER ML MODELS BASED ON BOT-IOT DATASET

| Model | Accuracy% | TPR% | FPR |
|---|---|---|---|
| MLP | 99.98 | 86.42 | 0.002 |
| CNN | 99.98 | 88.13 | 0.001 |
| GRU | 99.98 | 96.06 | 0.001 |
| LSTM | 99.99 | 94.69 | 0.0004 |
| Our model | 99.99 | 99.92 | 0.0003 |

TABLE IX. COMPARISON OF OUR PROPOSED MODEL WITH OTHER ML MODELS BASED ON TON_IOT DATASET

| Model | Accuracy% | TPR% | FPR |
|---|---|---|---|
| MLP | 99.67 | 99.51 | 0.03 |
| CNN | 99.88 | 99.75 | 0.01 |
| GRU | 97.85 | 96.95 | 0.27 |
| LSTM | 99.83 | 99.79 | 0.02 |
| Our model | 99.99 | 99.99 | 0.001 |

## VII. CONCLUSION

The smart city eases the way of life for the citizens by providing different services. However, it is vulnerable to various type of attacks due to the dependencies on ICT and the characteristics of the used technology which intimidate people to trust and move to smart city. A CTI can provide a secure smart city environment by monitoring the attacks in smart city and analyze the threat data to take prevention measures.

In this paper, we propose a hybrid DL model to classify threats. The proposed model uses CNN and QRNN models to improve the features extraction which increases the classification accuracy and lower the FPR. To improve the classification time to support real-time threats classification we use QRNN model. We use BoT-IoT and TON_IoT benchmark datasets to evaluate the proposed model. The results show the effectiveness of our model in improving the classification accuracy and lowering the FPR. In addition, the results show that the QRNN model can improve the classification time performance while providing high accuracy and low FPR as LSTM. Thus, the proposed model for CTI for smart city shows its' ability in analyzing and classifying the data accurately in real-time.

REFERENCES

[1] K. A. J. A. AlZaabi, *The Value of Intelligent Cybersecurity Strategies for Dubai Smart City*. Springer International Publishing, 2019.

[2] S. Tousley and S. Rhee, "Smart and Secure Cities and Communities," in *2018 IEEE International Science of Smart City Operations and Platforms Engineering in Partnership with Global City Teams Challenge (SCOPE-GCTC)*, 2018, pp. 7–11.

[3] J. Lee, J. Kim, and J. Seo, "Cyber attack scenarios on smart city and their ripple effects," in *2019 International Conference on Platform Technology and Service (PlatCon)*, 2019, pp. 1–5.

[4] P. Wu and H. Guo, "LuNet : A Deep Neural Network for Network Intrusion Detection," in *2019 IEEE Symposium Series on Computational Intelligence (SSCI)*, 2019, pp. 617–624.

[5] V. Behzadan and A. Munir, "Adversarial Exploitation of Emergent Dynamics in Smart Cities," in *2018 IEEE International Smart Cities Conference (ISC2)*, 2018, pp. 1–8.

[6] T. A. Butt and M. Afzaal, *Security and Privacy in Smart Cities: Issues and Current Solutions*. Springer International Publishing, 2019.

[7] L. Cui, G. Xie, Y. Qu, L. Gao, and Y. Yang, "Security and Privacy in Smart Cities : Challenges and Opportunities," *IEEE access*, vol. 6, pp. 46134–46145, 2018.

[8] A. Alibasic, R. Al Junaibi, Z. Aung, W. L. Woon, and M. A. Omar, "Cybersecurity for Smart Cities : A Brief Review," in *In International Workshop on Data Analytics for Renewable Energy Integration*, 2017, pp. 22–30.

[9] Z. A. Baig et al., "Future challenges for smart cities : Cyber-security and digital forensics," *Digit. Investig.*, vol. 22, pp. 3–13, 2017.

[10] T. Braun, B. C. M. Fung, F. Iqbal, and B. Shah, "Security and privacy challenges in smart cities," *Sustain. Cities Soc.*, vol. 39, pp. 499–507, 2018.

[11] H. Kettani and R. M. Cannistra, "On Cyber Threats to Smart Digital Environments," in *Proceedings of the 2nd International Conference on Smart Digital Environment. ACM*, 2018, pp. 183–188.

[12] N. Koroniotis, N. Moustafa, E. Sitnikova, and B. Turnbull, "Towards the Development of Realistic Botnet Dataset in the Internet of Things for Network Forensic Analytics: Bot-IoT Dataset," *Futur. Gener. Comput. Syst.*, vol. 100, pp. 779–796, 2018.

[13] F. Ahmad, A. Adnane, V. N. L. Franqueira, F. Kurugollu, and L. Liu, "Man-In-The-Middle Attacks in Vehicular Ad-Hoc Networks : Evaluating the Impact of Attackers ' Strategies," *Sensors*, pp. 1–19, 2018.

[14] G. Epiphaniou, A. Reviczky, P. Karadimas, H. Heidari, and S. Member, "Proactive Threat Detection for Connected Cars Using Recursive Bayesian Estimation," *IEEE Sens. J.*, vol. 18, no. 12, pp. 4822–4831, 2018.

[15] M. Liu, Z. Xue, X. He, and J. Chen, "Cyberthreat-Intelligence Information Sharing: Enhancing Collaborative Security," *IEEE Consumer Electronics Magazine*, vol. 8, no. April, IEEE, pp. 17–22, 2019.

[16] S. Abu, S. R. Selamat, A. Ariffin, and R. Yusof, "Cyber Threat Intelligence – Issue and Challenges," *Indones. J. Electr. Eng. Comput. Sci.*, vol. 10, no. 1, pp. 371–379, 2018.

[17] M. Conti, A. Dehghantanha, and T. Dargahi, *Cyberthreat Intelligence : Challenges and Opportunities*. Springer, 2018.

[18] K. Myat, N. Win, Y. Myo, and K. Khine, "Information Sharing of Cyber Threat Intelligence with their Issue and Challenges," *Int. J. Trend Sci. Res. Dev.*, vol. 3, no. 5, pp. 878–880, 2019.

[19] Y. Ghazi, Z. Anwar, R. Mumtaz, S. Saleem, and A. Tahir, "A Supervised Machine Learning Based Approach for Automatically Extracting High-Level Threat Intelligence from Unstructured Sources," in *2018 International Conference on Frontiers of Information Technology (FIT)*, 2018, pp. 129–134.

[20] L. Liu, O. De Vel, Q. Han, S. Member, and J. Zhang, "Detecting and Preventing Cyber Insider Threats : A Survey," *IEEE Commun. Surv. Tutorials*, vol. 20, no. 2, pp. 1397–1417, 2018.

[21] U. Noor, Z. Anwar, T. Amjad, and K. R. Choo, "A machine learning-based FinTech cyber threat attribution framework using high-level indicators of compromise," *Futur. Gener. Comput. Syst.*, vol. 96, pp. 227–242, 2019.

[22] N. Kaloudi and J. Li, "The AI-Based Cyber Threat Landscape : A Survey," *ACM Comput. Surv.*, vol. 53, no. 1, pp. 1–34, 2020.

[23] J. Lee, J. Kim, I. Kim, and K. Han, "Cyber Threat Detection Based on Artificial Neural Networks Using Event Profiles," *IEEE Access*, vol. 7, pp. 165607–165626, 2019.

[24] J. Abawajy, S. Huda, S. Sharmeen, and M. Mehedi, "Identifying cyber threats to mobile-IoT applications in edge computing paradigm," *Futur. Gener. Comput. Syst.*, vol. 89, pp. 525–538, 2018.

[25] X. Liang and T. Znati, "On the performance of intelligent techniques for intensive and stealthy DDos detection," *Comput. Networks*, vol. 164, p. 106906, 2019.

[26] A. A. Diro and N. Chilamkurti, "Distributed attack detection scheme using deep learning approach for Internet of Things," *Futur. Gener. Comput. Syst.*, vol. 82, pp. 761–768, 2018.

[27] D. S. Berman, A. L. Buczak, J. S. Chavis, and C. L. Corbett, "A Survey of Deep Learning Methods for Cyber Security," *Information*, vol. 10, no. 4, p. 122, 2019.

[28] A. Elsaeidy, K. S. Munasinghe, and D. Sharma, "A Machine Learning Approach for Intrusion Detection in Smart Cities," in *2019 IEEE 90th Vehicular Technology Conference (VTC2019-Fall)*, 2019, pp. 1–5.

[29] D. Li, L. Deng, M. Lee, and H. Wang, "IoT data feature extraction and intrusion detection system for smart cities based on deep migration learning," *Int. J. Inf. Manage.*, vol. 49, no. March, pp. 533–545, 2019.

[30] Y. N. Soe, Y. Feng, P. I. Santosa, R. Hartanto, and S. Kouichi, "Towards a Lightweight Detection System for Cyber Attacks in the IoT Environment Using Corresponding Features," *Electronics*, vol. 9, no. 1, pp. 1–19, 2020.

[31] M. Ge, X. Fu, N. Syed, Z. Baig, G. Teo, and A. Robles-kelly, "Deep Learning-based Intrusion Detection for IoT Networks," in *2019 IEEE 24th Pacific Rim International Symposium on Dependable Computing (PRDC)*, 2019, pp. 256–25609.

[32] J. Alsamiri and K. Alsubhi, "Internet of Things Cyber Attacks Detection using Machine Learning," *Int. J. Adv. Comput. Sci. Appl.*, vol. 10, no. 12, 2019.

[33] A. Khraisat, I. Gondal, P. Vamplew, J. Kamruzzaman, and A. Alazab, "A Novel Ensemble of Hybrid Intrusion Detection System for Detecting Internet of Things Attacks," *Electronics*, vol. 8, no. 11, p. 1210, 2019.

[34] I. Ullah and Q. H. Mahmoud, "A Two-Level Flow-Based Anomalous Activity Detection System for IoT Networks," *Electronics*, vol. 9, no. 3, p. 530, 2020.

[35] P. Sornsuwit and S. Jaiyen, "A New Hybrid Machine Learning for Cybersecurity Threat Detection Based on Adaptive Boosting," *Appl. Artif. Intell.*, vol. 33, no. 5, pp. 462–482, 2019.

[36] V. L. L. Thing, "IEEE 802.11 Network Anomaly Detection and Attack Classification: A Deep Learning Approach," in *In 2017 IEEE Wireless Communications and Networking Conference (WCNC)*, 2017, pp. 1–6.

[37] S. Garg et al., "A Hybrid Deep Learning-Based Model for Anomaly

[37] (continued) Detection in Cloud Datacenter Networks," *IEEE Trans. Netw. Serv. Manag.*, vol. 16, no. 3, pp. 924–935, 2019.

[38] M. Mehedi, A. Gumaei, A. Alsanad, M. Alrubaian, and G. Fortino, "A hybrid deep learning model for efficient intrusion detection in big data environment," *Inf. Sci. (Ny).*, vol. 513, pp. 386–396, 2020.

[39] R. Vinayakumar, S. Kp, and P. Poornachandran, "Applying Convolutional Neural Network for Network Intrusion Detection," in *In 2017 International Conference on Advances in Computing, Communications and Informatics (ICACCI)*, 2017, pp. 1222–1228.

[40] W. E. I. Wang *et al.*, "HAST-IDS : Learning Hierarchical Spatial-Temporal Features Using Deep Neural Networks to Improve Intrusion Detection," *IEEE Access*, vol. 6, pp. 1792–1806, 2018.

[41] Y. Lecun, Y. Bengio, and G. Hinton, "Deep learning," *Nature*, vol. 521, no. 7553, pp. 436–444, 2015.

[42] M. Amine, L. Maglaras, S. Moschoyiannis, and H. Janicke, "Deep learning for cyber security intrusion detection : Approaches , datasets , and comparative study," *J. Inf. Secur. Appl.*, vol. 50, p. 102419, 2020.

[43] N. Hasan, R. N. Toma, A. Nahid, M. M. M. Islam, and J. Kim, "Electricity Theft Detection in Smart Grid Systems : A CNN-LSTM Based Approach," *Electr. Th. Detect. Smart Grid Syst. A CNN-LSTM Based Approach*, vol. 12, no. 17, p. 3310, 2019.

[44] D. Kwon, K. Natarajan, S. C. Suh, H. Kim, and J. Kim, "An Empirical Study on Network Anomaly Detection using Convolutional Neural Networks," in *In 2018 IEEE 38th International Conference on Distributed Computing Systems (ICDCS)*, 2018, pp. 1595–1598.

[45] H. Liu, B. Lang, M. Liu, and H. Yan, "Knowledge-Based Systems CNN and RNN based payload classification methods for attack detection," *Knowledge-Based Syst.*, vol. 163, pp. 332–341, 2019.

[46] A. Khan, "RNN-LSTM-GRU based language transformation," *Soft Comput.*, vol. 23, no. 24, pp. 13007–13024, 2019.

[47] F. Bolelli, L. Baraldi, F. Pollastri, and C. Grana, "A Hierarchical Quasi-Recurrent approach to Video Captioning," in *2018 IEEE International Conference on Image Processing, Applications and Systems (IPAS)*, 2018, pp. 162–167.

[48] S. Merity, C. Xiong, and R. Socher, "Quasi-Recurrent Neural Network," in *arXiv*, 2017, pp. 1–11.

[49] M. Wang *et al.*, "Quasi-fully Convolutional Neural Network with Variational Inference for Speech Synthesis," in *ICASSP 2019-2019 IEEE International Conference on Acoustics, Speech and Signal Processing (ICASSP)*, 2019, pp. 7060–7064.

[50] J. Huang and Y. Feng, "Optimization of Recurrent Neural Networks on Natural Language Processing," in *Proceedings of the 2019 8th International Conference on Computing and Pattern Recognition*, 2019, pp. 39–45.

[51] P. Wu, H. Guo, and N. Moustafa, "Pelican : A Deep Residual Network for Network Intrusion Detection," *arXiv*, vol. 2001.08523, 2020.

[52] D. Yao, M. Wen, X. Liang, Z. Fu, K. Zhang, and B. Yang, "Energy Theft Detection With Energy Privacy Preservation in the Smart Grid," *IEEE Internet Things J.*, vol. 6, no. 5, pp. 7659–7669, 2019.

[53] N. Moustafa, "TON_IoT Datasets," 2020. .

[54] H. Safa, M. Nassar, and W. A. R. Al Orabi, "Benchmarking Convolutional and Recurrent Neural Networks for Malware Classification," in *2019 15th International Wireless Communications & Mobile Computing Conference (IWCMC)*, 2019, pp. 561–566.